\documentclass[prd,eqsecnum,twocolumn,amsfonts,showpacs,amssymb]{revtex4}
\usepackage{graphicx}
\graphicspath{{./figures/}}

\usepackage{bm}

\newcommand{\beq}{\begin{equation}}
\newcommand{\eeq}{\end{equation}}
\newcommand{\beqs}{\begin{eqnarray}}
\newcommand{\eeqs}{\end{eqnarray}}
\newcommand{\lsim}{\mathrel{\raisebox{-
.6ex}{$\stackrel{\textstyle<}{\sim}$}}}

\newcommand{\dslash}{\partial\hspace{-0.07in}\slash}

\newcommand{\drawsquare}[2]{\hbox{%
\rule{#2pt}{#1pt}\hskip-#2pt
\rule{#1pt}{#2pt}\hskip-#1pt
\rule[#1pt]{#1pt}{#2pt}}\rule[#1pt]{#2pt}{#2pt}\hskip-#2pt
\rule{#2pt}{#1pt}}
\newcommand{\fund}{\raisebox{-.5pt}{\drawsquare{6.5}{0.4}}}

\begin{document}

\title{Renormalization-Group Flows and Fixed Points in Yukawa Theories}

\author{Esben M{\o}lgaard$^{a,b}$ and Robert Shrock$^b$}

\affiliation{(a) \ The Centre for Cosmology and Particle Physics Phenomenology CP$^3$-Origins, and the Danish Institute for Advanced Study DIAS,\\
University of Southern Denmark, Campusvej 55, DK-5230 Odense M, Denmark}

\affiliation{(b) \ C. N. Yang Institute for Theoretical Physics and 
Department of Physics and Astronomy \\
Stony Brook University, Stony Brook, NY 11794, USA }

\begin{abstract}

We study renormalization-group flows in Yukawa theories with massless fermions,
including determination of fixed points and curves that separate regions of
different flow behavior.  We assess the reliability of perturbative
calculations for various values of Yukawa coupling $y$ and quartic scalar
coupling $\lambda$ by comparing the properties of flows obtained with the beta
functions of these couplings calculated to different orders in the loop
expansion. The results provide a determination of the region in $y$ and
$\lambda$ where calculations up to two loops can yield reasonably reliable
results.

\vspace{0.5cm}
\noindent

{ \footnotesize  \it Preprint: CP3-Origins-2014-008 DNRF90, DIAS-2014-8; 
YITP-Stony Brook-2013-39}
\end{abstract}


\maketitle


\section{Introduction}

The dependence of the coupling constants in a quantum field theory on the
Euclidean momentum scale $\mu$, at which they are measured is of fundamental
importance.  This behavior is described by the beta functions for these
couplings \cite{rg}.  In a theory with two or more couplings, a change in $\mu$
thus induces a renormalization-group (RG) flow in the space of couplings. The
RG flow typically involves some infrared (IR) or ultraviolet (UV) fixed points,
and one can characterize these as being attractive or repulsive along certain
directions in the space of couplings.  If the couplings are sufficiently small,
then the respective beta functions can be reliably calculated perturbatively.
As one or more of these couplings increases in magnitude, higher-loop
contributions to the various beta functions become important, motivating
calculations of these beta functions to higher loop order to obtain reliable
results for RG flows (trajectories) and fixed points.  If one or more couplings
becomes too large, then it may not be possible to describe the RG flows, or,
more generally, the properties of the theory, using perturbative calculations.

A general criterion for the reliability of a perturbative calculation is that
if one calculates some quantity to a given loop order, then there should not be
a large fractional change in this quantity if one computes it to one higher
order in the loop expansion. Thus, in a situation where a putative fixed point
occurs at moderately strong coupling, it is important to study how the value of
the coupling(s) at this fixed point change(s) if one calculates the beta
function(s) to higher loop order.  For example, an asymptotically free
non-Abelian gauge theory with sufficiently many fermions in a given
representation has an IR fixed point (IRFP) \cite{b2}. If the number of
fermions is only slightly less than the maximum allowed by the constraint of
asymptotic freedom, this IRFP occurs at weak coupling \cite{bz}.  As the number
of fermions is decreased, the IRFP moves to stronger coupling, and studies have
been carried out of the effect of higher-loop terms in the beta function of the
gauge coupling in this case \cite{bvh}.  One may also investigate a possible
ultraviolet fixed point (UVFP) in an infrared-free theory such as U(1) gauge
theory with higher-loop calculations (e.g., \cite{holdom,uvfp} and references
therein).

It is also of considerable interest to investigate renormalization-group flows
in the more complicated case of quantum field theories that depend on more than
one interaction coupling. There have been many studies of such flows for
theories and ranges of momentum scale $\mu$ where the couplings are reasonably
weak, so that perturbative calculations are reasonably accurate. This is the
case for computations of RG flows of the SU(3)$_c$, SU(2)$_L$, and U(1)$_Y$
gauge couplings in the Standard Model (SM) or the Minimal Supersymmetric
Standard Model (MSSM) upward from a reference scale of, say, 1 TeV, up to
higher scales such as $10^{16}$ GeV. There has also been interest in
calculating the RG flow of the elements of Yukawa matrices in the SM and MSSM,
and the quartic Higgs coupling $\lambda_{SM}$ in the SM, from the 1 TeV scale
to higher scales.  Again, these RG flows can be reasonably well described by
perturbative calculations, although with the measured value of the Higgs-like
boson observed by the LHC, $m_H \simeq 126$ GeV (whence in the SM,
$\lambda_{SM}(\mu) \simeq 0.13$ at $\mu=m_H$), in the absence of new physics
effects at intermediate scales, it follows that $\lambda_{SM}(\mu)$ would 
decrease through zero at a high scale $\mu \sim 10^{10 \pm 1}$ GeV, implying
that the SM, by itself, would be metastable above this scale
\cite{metastability}-\cite{agkms}.

In this paper we study renormalization-group flows in Yukawa theories and
assess the reliability of perturbative calculations of these flows for a
substantial range of Yukawa and quartic scalar couplings. The method that we
use for this purpose is to compare the properties of flows that we obtain with
the beta functions of these couplings calculated to different orders in the
loop expansion. In order to focus on the essential features in as simple a
framework as possible, we study scalar-fermion models without any gauge fields.
We construct these models so that the global symmetries forbid any Dirac or
Majorana fermion mass terms, and we also consider the limit where scalar masses
are negligibly small relative to the scales $\mu$ of interest.  These models
depend on two dimensionless couplings, a quartic self-coupling $\lambda$ for
the scalar field and a Yukawa coupling $y$.  The beta functions for these
couplings comprise a set of coupled first-order ordinary differential equations
describing how the couplings vary as functions of $\mu$. Integrating this set
of differential equations, we determine their renormalization-group flows as
functions of $\mu$.  To do this, we choose an initial scale, $\mu_0$, where the
magnitudes of the couplings are sufficiently small that perturbative
calculations may be reliable, and then perform the integration.  Our method is
to compare RG flows calculated using different loop orders for the two beta
functions.  We recall the basic fact that in these theories the quartic scalar
self-coupling $\lambda$ must be positive for the boundedness of the energy and
equivalently the stability of the theory.  As will be evident in our results,
RG flows may take a theory with positive $\lambda$ to one with negative
$\lambda$.  In this case, two comments are necessary.  Strictly speaking, for a
sufficiently small range of negative $\lambda$ the theory may still be
metastable, with a sufficiently long tunneling time that our perturbative
calculations may be physically meaningful.  However, for negative values of
$\lambda$ of sufficiently large magnitude, the theory is simply unstable, and
the perturbative analysis is not applicable or meaningful.  In most of our
analytic discussions, therefore, we will implicitly take $\lambda$ to be
positive.

We remark on some earlier related work on Yukawa models. As is well known,
Yukawa proposed such models \cite{yukawa1935} as an approach to understanding
the binding of nucleons in nuclei, and pion exchange between nucleons does,
indeed, play an important role in this binding.  Of course, the physics here
involves the exchange of a light approximate Nambu-Goldstone boson between two
baryons, with the baryons being much heavier than the exchanged $\pi$ meson, as
indicated by the ratio of masses $m_\pi/m_N = 0.15$.  This is quite different
from our our models, for which, by construction, a global chiral symmetry
forbids any fermion mass fermions and the scalar mass is taken to be negligibly
small relative to the interval of Euclidean momentum scales $\mu$ for which we
integrate the beta functions to calculate the RG flows.  Some early studies of
perturbative RG equations for Standard Model Yukawa couplings included
Refs. \cite{cel,earlyfm}.  It was recognized early on that the one-loop beta
function for a scalar theory without fermions is positive, this theory is,
perturbatively, at least, IR-free; that is, as $\mu \to 0$, $\lambda(\mu) \to
0$.  However, it was also recognized that if one adds fermions to this scalar
theory to get a full scalar-fermion Yukawa theory, then the fermions contribute
a negative term proportional to $y^4$ in the beta function $d\lambda/d\ln \mu$,
and hence, for sufficiently large $y$, this can reverse the sign of the full
one-loop term in this beta function and hence possibly render the scalar
coupling in the Yukawa theory nontrivial \cite{earlyfm}. This motivated fully
nonperturbative studies, and these were carried out using lattice studies with
dynamical fermions \cite{yuk} (some recent work includes \cite{yukrecent}).
One may obtain a Yukawa theory starting from a full gauge-fermion-Higgs theory
by turning off the gauge couplings.  In this framework, a natural approach is
to start with a chiral gauge theory (exemplified by the Standard Model), which
forbids bare fermion masses in the Lagrangian.  However, owing to fermion
doubling on the lattice, it has been challenging to implement chiral gauge
theories on the lattice.  We believe, therefore, that there is continuing
interest in pursuing analyses of renormalization-group evolution of continuum
Yukawa theories using perturbatively calculated beta functions.  Indeed, simple
scalar-fermion models have been of recent interest in studies of quasi-scale
invariant behavior (e.g., \cite{sfm}; see also \cite{rg2,agkms}).

This paper is organized as follows. In Sect. II we define our notation for the
relevant variables and beta functions.  In Sect. III we study a scalar-fermion
model with an ${\rm SU}(2) \otimes {\rm U}(1)$ global symmetry group.  In
Sect. IV we generalize this analysis to a model with $N_f$ copies (``flavors'')
of fermions and an ${\rm SU}(N) \otimes {\rm SU}(N_f) \otimes {\rm U}(1)$ 
global symmetry group. Our conclusions are contained in Sect. V. 


\section{Beta Functions}\label{sec:betaFuncs}

The beta functions describing the dependence of the running couplings
$y=y(\mu)$ and $\lambda = \lambda(\mu)$ on the scale $\mu$ where they are
measured are
\beq
\beta_y \equiv \frac{dy}{dt} \ , \quad \quad 
\beta_\lambda \equiv \frac{d\lambda}{dt} \ , 
\label{betaybetalam}
\eeq
where $dt = d\ln (\mu/\mu_0)$, where $\mu_0$ is an initial value of the
reference scale.  (The $\mu$ dependence of $y$ and $\lambda$ is
implicitly understood below but the argument will often be suppressed in the
notation.) These beta functions can be expressed as a sum of $\ell$-loop terms
as
\beq
\beta_y = \sum_{\ell=1}^\infty \frac{ b_y^{(\ell)}}{(4\pi)^{2\ell}} \ , 
\quad \quad 
\beta_\lambda = \sum_{\ell=1}^\infty \frac{ b_\lambda^{(\ell)}}{(4\pi)^{2\ell}}
\ , 
\label{betaseries}
\eeq
where $b_y^{(\ell)}/(4\pi)^{2\ell}$ and 
where $b_\lambda^{(\ell)}/(4\pi)^{2\ell}$ denote the $\ell$-loop contributions
to $\beta_y$ and $\beta_\lambda$, respectively. 

It will also be convenient to define the variables 
\beq
a_y \equiv \frac{y^2}{(4\pi)^2} \ , \quad \quad 
a_\lambda \equiv \frac{\lambda}{(4\pi)^2} \ , 
\label{ay_alam}
\eeq
which will be used for the ${\rm SU}(2) \otimes {\rm U}(1)$ model 
studied below. For the ${\rm SU}(N) \otimes {\rm SU}(N_f) \otimes {\rm U}(1)$ 
model and, in particular, for the limit (\ref{lnn}), we define 
the variables 
\beq
\bar a_y \equiv \frac{y^2 N}{(4\pi)^2} \ , \quad \quad 
\bar a_\lambda \equiv \frac{\lambda N}{(4\pi)^2} \ .
\label{aybaralambar}
\eeq

Correspondingly, for the ${\rm SU}(2) \otimes {\rm U}(1)$ model we define
the beta functions 
\beq
\beta_{a_y} \equiv \frac{da_y}{dt} = \frac{2y}{(4\pi)^2} \beta_y\ , \quad \quad 
\beta_{a_\lambda} \equiv \frac{da_\lambda}{dt} = \frac{1}{(4\pi)^2}
  \beta_{a_\lambda} \ , 
\label{beta_ayalam}
\eeq
with the series expansions 
\beq
\beta_{a_y} = 
\sum_{\ell=1}^\infty b_{a_y}^{(\ell)} \ , 
\quad \quad 
\beta_{a_\lambda} = 
\sum_{\ell=1}^\infty b_{a_\lambda}^{(\ell)} \ . 
\label{betaseriesa}
\eeq
From the relations above, it follows that
\beq
b_{a_y}^{(\ell)} = \frac{2y}{(4\pi)^{2(\ell+1)}} \, b_y^{(\ell)} \ , 
\quad\quad 
b_{a_\lambda}^{(\ell)} = \frac{1}{(4\pi)^{2(\ell+1)}} \, b_\lambda^{(\ell)} 
\label{bylam_bayalam_ell_rel}
\eeq
We denote the $n$-loop ($n\ell$) beta functions as $\beta_{a_y,n \ell}$ and 
$\beta_{a_\lambda,n \ell}$. 

Similarly, for the ${\rm SU}(N) \otimes {\rm SU}(N_f) \otimes {\rm U}(1)$ 
model, we define the beta functions 
\beq
\beta_{\bar a_y} \equiv \frac{d\bar a_y}{dt} = \frac{2yN}{(4\pi)^2} \, \beta_y 
\label{beta_ay_betay_rel}
\eeq
and
\beq
\beta_{\bar a_\lambda} \equiv \frac{d\bar a_\lambda}{dt} = 
\frac{N}{(4\pi)^2} \, \beta_\lambda
\label{beta_alam_betalam_rel}
\eeq
with series expansions analogous to those in Eq. (\ref{betaseriesa}) with $a_y$
and $a_\lambda$ replaced by $\bar a_y$ and $\bar a_\lambda$, respectively.  In
the latter case, the LNN limit (\ref{lnn}) will generally be understood.

As discussed in the introduction, these beta functions form a set of two
coupled differential equations.  We integrate these for each of the two models
that we study to calculate the resultant RG flows.  A point in the
multidimensional space of couplings where all of the beta functions vanish
simultaneously is, formally, a renormalization-group fixed point (FP). In
general, RG flows may include the presence of one or more ultraviolet (UV)
fixed point(s) if the beta functions vanish as $\mu \to \infty$ and/or 
infrared (IR) fixed point(s), where the beta functions vanish as $\mu \to 0$.
In general, a fixed point may be stable along some directions and unstable
along others.  If the particle content of the theory does not change
along the RG flow from $\mu_0$ to the fixed point, then it is an exact UV or IR
fixed point.  In the vicinity of a (formal) fixed point, the RG flows are
slow, so that the theories exhibit approximate scale-invariance.

For our comparative study we will perform the integrations to calculate the RG
flows with the beta functions $\beta_{a_y}$ and $\beta_{a_\lambda}$ calculated
to various different loop orders. We denote these as follows. For the 
${\rm SU}(2) \otimes {\rm U}(1)$ model, the 
calculation using the $\beta_{a_y,n\ell}$ and $\beta_{a_\lambda,k\ell}$
beta functions is denoted $(n,k)$.  The specific cases 
for which we perform the integrations are 
\begin{itemize}

\item  (1,1), i.e., $\beta_{a_y,1\ell}$ and $\beta_{a_\lambda,1\ell}$

\item  (1,2), i.e., $\beta_{a_y,1\ell}$ and $\beta_{a_\lambda,2\ell}$

\item  (2,1), i.e., $\beta_{a_y,2\ell}$ and $\beta_{a_\lambda,1\ell}$

\item  (2,2), i.e., $\beta_{a_y,2\ell}$ and $\beta_{a_\lambda,2\ell}$

\end{itemize} 

We use the same notation to describe the four cases for the ${\rm SU}(N)
\otimes {\rm SU}(N_f) \otimes {\rm U}(1)$ model, so that in this context, the
case (1,1) refers to an RG calculation using $\beta_{\bar a_y,1\ell}$ and
$\beta_{\bar a_\lambda,1\ell}$ and so forth for the other cases.  Some remarks
are in order here.  For a perturbative calculation of quantities in a theory
with multiple couplings, a general procedure would be to calculate to similar
orders in the various couplings if they are equally large and significant for
the physics, and to calculate to higher order in a coupling that is larger.
Thus, for example, in a Standard-Model process, one may only need to calculate
to lowest order in electroweak couplings, but to higher order in the QCD
coupling.  Such a calculation is consistent in the sense that one has included
higher-order terms in a larger coupling. Ref. \cite{agkms} obtained the result
that Weyl consistency conditions are maintained only if one uses the beta
functions $\beta_{a_g,(n+2)\ell}$, $\beta_{a_y,(n+1)\ell}$, and
$\beta_{a_\lambda,n \ell}$, where $g$ denotes a gauge coupling and $a_g \equiv
g^2/(4\pi)^2 = \alpha/(4\pi)$ (see also \cite{gss}).

In this type of study there are several obvious caveats.  First, clearly, as
couplings increase in strength, perturbative calculations become progressively
less reliable. This is, indeed, a motivation for our present work - to assess
quantitatively where this reduction in reliability occurs in the case of
scalar-fermion models depending on two coupling constants.  Second, higher-loop
terms in beta functions of multi-coupling theories are generically
scheme-dependent, and the positions of fixed points are hence also
scheme-dependent. Indeed, scheme dependence is also present in higher-loop
calculations in quantum chromodynamics (QCD). As in common practice in QCD, we
use results computed with the $\overline{MS}$ scheme \cite{msbar}.  One can
assess the effect of scheme dependence of RG flows and fixed points by
comparing these in different schemes \cite{bvh}.  However, many scheme
transformations that are acceptable in the vicinity of a fixed point at zero
coupling (e.g., a UVFP in an asymptotically free gauge theory, or an IRFP in an
infrared-free theory) are not acceptable at a fixed point that occurs at a
moderately strong coupling, because they produce various unphysical pathologies
\cite{uvfp,sch,ta}. A third caveat, related to the first, is that if one or
more of the couplings is (are) sufficiently large, the Yukawa and/or quartic
scalar self-interaction may lead to nonperturbative phenomena such the
formation of a fermion condensate, a vacuum expectation value (VEV) for the
scalar field, and/or fermion-fermion bound states (see, e.g., \cite{rg2},
\cite{boundstates}).  In the case where the coefficient of the quadratic term
in the scalar potential $V$ is zero, there is the related possibility of a
nonperturbative generation of a nonpolynomial term in $V$, whose minimum could
lead to a nonzero VEV for the scalar field \cite{cw}. Early studies of the
stability of a theory in the presence of this phenomenon and associated related
bounds on fermion and Higgs masses include \cite{earlyfm,
fermionhiggsstability}.

If fermion condensation occurs at some scale $\mu_c$
(where the subscript $c$ for condensate) in the vicinity of a formal IR fixed
point, then the originally massless fermions gain dynamical masses,
spontaneously breaking the approximate scale invariance in the theory near to
an apparent RG fixed point.  In the low-energy effective field theory
applicable for scales $\mu < \mu_c$, one integrates these fermions out, thereby
obtaining different beta functions.  Thus, in this case, the formal fixed point
would only be approximate rather than exact, since after the fermion
condensation, the beta functions and flows would be different.  This
spontaneous symmetry breaking (SSB) of the approximate scale invariance
generically leads to the appearance of a corresponding Nambu-Goldstone boson,
the dilaton.  This dilaton is not massless, since the beta functions in the
vicinity of the fixed point were small but not precisely zero.

If $\mu_\phi^2 < 0$ so that there is a VEV for the scalar field, then the
Yukawa coupling leads to a mass for the fermion field(s) of the form $m_f
\propto yv$.  However, since the VEV $v = (-\mu_\phi^2/\lambda)^{1/2}$ and
since we assume that $|\mu_\phi|$ is much smaller than the reference scales
$\mu$ over which we integrate the renormalization-group equations, it follows
that for moderate values of the ratio $y^2/\lambda$, the resultant fermion
masses $m_f \propto y(-\mu_\phi^2/\lambda)^{1/2}$ are negligible relative to
the interval of $\mu$ that we study.


\section{${\rm SU}(2) \otimes {\rm U}(1)$ Model} 


\subsection{Field Content and Symmetry Group} 

The first model that we study is motivated by the leptonic sector of the
Standard Model, with the gauge interactions turned off.  It includes a fermion
$\psi^a_L$ which is a doublet under SU(2) with weak hypercharge $Y_\psi$
and a $\chi_R$, which is a singlet under SU(2) with weak hypercharge
$Y_\chi$, together with the usual scalar field $\phi^a$ transforming as a
doublet under SU(2) with weak hypercharge $Y_\phi$.  Here, $a=1,2$ is an
SU(2) group index which will often be suppressed in the notation.  We
assume that these hypercharges are nonzero and that $Y_\psi \ne Y_\chi$. Since
we have set the gauge couplings to zero, the ${\rm SU}(2) \otimes {\rm
U}(1)$ is a global symmetry group.  As in the Standard Model, we set
\beq
Y_\phi = Y_\psi - Y_\chi
\label{hypercharge_rel}
\eeq
to ensure that a the Yukawa interaction term is invariant under the global
symmetry.  The Lagrangian for this model is
\beqs
{\cal L} & = & \bar\psi_L i \dslash \psi_L + \bar\chi_R i \dslash \chi_R 
- [ y \bar\psi_L \chi_R \phi + h.c. ] \cr\cr
& + & \partial_\mu \phi^\dagger \, \partial^\mu \phi - \mu_\phi^2 \phi^\dagger
\phi - \lambda (\phi^\dagger \phi)^2 \ .
\label{lagrangian1}
\eeqs
Without loss of generality, we can make $y(\mu_0)$ real and positive at a given
value $\mu_0$ (by changing the phase of $\psi_L$ or $\chi_R$ or $\phi$).
We assume that this is done. We allow $\mu_\phi^2$ of either sign but assume that
$|\mu_\phi^2|$ is negligibly small compared with the range of $\mu^2$ of
interest for our study of RG flows \cite{mphi} (see also the end of Section
\ref{sec:betaFuncs}). The global SU(2) symmetry forbids the Majorana bilinear
$\psi^{a, \ T}_L C \psi^b_L$ and the Dirac bilinear $\bar\psi_{a,L} \chi_R$
from occurring in ${\cal L}$.  Since $Y_\chi$ is taken to be nonzero, the U(1)
symmetry forbids the Majorana bilinear $\chi_R^T C \chi_R$ (as well as $\psi^{a
\ T}_L C \psi^b_L$ and $\bar\psi_{a,L} \chi_R$ bilinears). Thus, the condition
that ${\cal L}$ be invariant under this global symmetry group implies that the
fermions are massless.


\subsection{Beta Functions}

The one-loop and two-loop coefficients in the beta functions 
$\beta_y$ and $\beta_\lambda$ can be extracted, with 
the requisite changes to match our normalizations, from previous 
calculations (which were done in the $\overline{MS}$ scheme) 
\cite{cel,betastability,betafn}.  They are 
\beq
b_y^{(1)} = \frac{5}{2}y^3 
\label{by_1loop}
\eeq
\beq
b_y^{(2)} = 3y( -y^4 -4y^2\lambda + 2\lambda^2 ) 
\label{by_2loop}
\eeq
\beq
b_\lambda^{(1)} = 2(12\lambda^2 + 2y^2\lambda-y^4)
\label{blam_1loop}
\eeq
\beq
b_\lambda^{(2)} = -312\lambda^3-48y^2\lambda^2-y^4\lambda+10y^6 \ . 
\label{blam_2loop}
\eeq
In terms of the variables $a_y$ and $a_\lambda$ used for the figures, 
\beq
b_{a_y}^{(1)} = 5 a_y^2 
\label{bay_1loop}
\eeq
\beq
b_{a_y}^{(2)} = 6a_y( -a_y^2 -4a_ya_\lambda + 2a_\lambda^2 ) 
\label{bay_2loop}
\eeq
\beq
b_{a_\lambda}^{(1)} = 2(12a_\lambda^2 + 2a_ya_\lambda-a_y^2)
\label{balam_1loop}
\eeq
\beq
b_{a_\lambda}^{(2)} = -312a_\lambda^3-48a_y a_\lambda^2-a_y^2a_\lambda+10a_y^3 
\ . 
\label{balam_2loop}
\eeq

We comment on some properties of $\beta_y$ or equivalently, $\beta_{a_y}$. 
We recall that at the initial point $\mu_0$ where we start our integrations of
the renormalization group equations, we have, with no loss of generality,
rendered $y$ real and positive. A first comment is that because $\beta_y$ has
an overall factor of $y$, and $\beta_{a_y}$ has an overall factor of $a_y$, 
it follows that the flow in $y$ can never take $y$ through zero to negative
values of $y$, and the flow in $a_y$ can never take $a_y$ through zero to 
negative values of $a_y$.  

The fact that $b_{a_y}^{(1)} > 0$ means that for sufficiently small $a_y$ and
$a_\lambda$, $\beta_{a_y} > 0$, i.e., as $\mu$ decreases from the UV to the 
IR, the Yukawa coupling $y$ decreases.  At the two-loop level,
\beq
b_{a_y}^{(2)} > 0 \quad {\rm if} \quad a_\lambda > ( 1 + \sqrt{3/2} \ )a_y 
= 2.2247a_y \ , 
\label{by_2loop_pos_condition}
\eeq
to the given floating-point accuracy.  If these conditions are satisfied, then
the two-loop coefficient contributes to $\beta_{a_y}$ with the same sign as the
one-loop coefficient and increases the rate of change of $a_y$ as a
function of $\mu$.  If, on the other hand $a_\lambda < (1+\sqrt{3/2} \
)a_y$, then $b_{a_y}^{(2)} < 0$, so $b_{a_y}^{(2)}$ contributes to
$\beta_{a_y}$ with a sign opposite to that of $b_{a_y}^{(1)}$.  In this case,
it is possible for $\beta_{a_y}$ to vanish at the two-loop level.  The
condition for this to happen is that either $a_y=0$ for some $\mu$ or
(again suppressing the argument, $\mu$) that 
\beq
5a_y + 6(-a_y^2-4a_ya_\lambda+2a_\lambda^2) = 0 \ . 
\label{betay_2loop_zero_eq}
\eeq
Solving this equation for $a_y$ yields the physical solution 
\beq
a_y = \frac{5}{12} - 2a_\lambda + \frac{1}{12} \sqrt{864a_\lambda^2 - 
240a_\lambda + 25} \ . 
\label{beta_2loop_zero_aysol}
\eeq
(The polynomial in the square root is positive-definite.) 
Equivalently, solving Eq. (\ref{betay_2loop_zero_eq}) for $a_\lambda$ yields
\beq
a_\lambda = a_y + \frac{1}{6}\sqrt{3a_y(18a_y-5)}  \ , 
\label{beta_2loop_zero_alamsol}
\eeq
which is physical if $a_y \ge 5/18$, i.e., $y \ge (4\pi/3)\sqrt{5/2} =
6.623$. Evidently, this zero of $\beta_{a_y,2\ell}$ is only possible for such
large values of $y$ that one must anticipate significant corrections from
higher-loop terms in $\beta_{a_y}$. In passing, we note that the other solution
of Eq. (\ref{betay_2loop_zero_eq}) for $\lambda$ with a minus sign in front of
the square root is unphysical, since it can lead to a negative
$\lambda$. (As noted before, we do not attempt to consider a metastable
situation with a negative $\lambda$ of small magnitude.)  
Also, the other solution of Eq. (\ref{betay_2loop_zero_eq}) for
$a_y$ with a minus sign in front of the square root in
Eq. (\ref{beta_2loop_zero_aysol}) is unphysical because it can lead
to a value of $a_y < 5/18$.  Setting $a_y=5/18$ in
Eq. (\ref{beta_2loop_zero_alamsol}) yields $a_\lambda=a_y=5/18$, and similarly,
setting $a_\lambda=5/18$ in Eq. (\ref{beta_2loop_zero_aysol}) yields
$a_y=a_\lambda=5/18$.

We next remark on some properties of $\beta_{a_\lambda}$. We find that 
\beq
b_{a_\lambda}^{(1)} = 0 \quad {\rm if} \quad a_\lambda = 
\frac{(\sqrt{13} - 1)}{12} \, a_y =0.21713a_y  
\label{blam_1loop_zero_ysol} 
\eeq
and
\beq
b_{a_\lambda}^{(1)} > 0 \quad {\rm if} \quad a_\lambda > 
\frac{(\sqrt{13} - 1)}{12} \, a_y \ , 
\label{blam_1loop_pos_condition}
\eeq
or equivalently, $a_y < (1+\sqrt{13} \ )a_\lambda = 4.60555 a_\lambda$. 
The condition that $b_{a_\lambda}^{(2)}=0$ is a cubic
equation in $a_\lambda$ and separately a cubic equation in $a_y$. 
We find that if $a_y=(1+\sqrt{13} \ )a_\lambda$, so that
$b_\lambda^{(1)}=0$, then 
\beq
b_{a_\lambda}^{(2)} = \frac{2(13+55\sqrt{13})}{(4\pi)^6} \,
\lambda^3 = (1.073 \times 10^{-4}) \lambda^3 \ . 
\label{blam_2loop_value_at_blam_1loopzero}
\eeq
Hence, if the values of $a_\lambda$ and $a_y \ne 0$ are such that the one-loop
contribution to $\beta_{a_\lambda} = 0$, then at the two-loop level, 
$\beta_{a_\lambda} > 0$. 

In the special case where $a_y=0$, we find that if we consider
$\beta_{a_\lambda,2\ell}$, a non-trivial fixed point appears at
\beq
  a_\lambda^*=\frac{1}{13} = 0.076923\ .
\label{ay0alamfp}
\eeq
This fixed point is repulsive in the $a_y$-direction, since for lower values of
$a_\lambda$ (while keeping $a_y=0$), $b^{(1)}_{a_\lambda}$ drives the flow down, and
for higher, $b^{(2)}_{a_\lambda}$ drives it up.

We next give some illustrative numerical evaluations.  Let us consider that the
theory is such that at some reference scale $\mu_0$, $y(\mu_0)$ and
$\lambda(\mu_0)$ have the values $y(\mu_0)=1$ and $\lambda(\mu_0)=1$.  If one
were to consider turning on gauge fields (and adding quarks so that this theory
is free of gauge anomalies), then these would be rather large physical values
of these couplings.  For reference, considering only the third generation in
the Standard Model (SM) and using the relation for a fermion mass in terms of
the Yukawa coupling and the Higgs vacuum expectation value, $\langle \phi
\rangle_0$, namely
\beq
y_f \langle \phi \rangle_0 = y_f \frac{v}{\sqrt{2}} = m_f \ , 
\label{yfsm}
\eeq
where $v=246$ GeV, one has the rough values $y_\tau \simeq 1 \times 10^{-2}$,
$y_b \simeq 2 \times 10^{-2}$, and $y_t \simeq 1$. Further, using the relation
for the Higgs boson mass $m_H$ in the Standard Model, namely,
\beq
m_H = (2\lambda)^{1/2} v
\label{mh}
\eeq
one has $\lambda(\mu) = 0.13$ at $\mu=m_H=126$ GeV, as noted above.  So the
illustrative reference values $y(\mu_0)=\lambda(\mu_0) = 1$ that we have taken
may be considered to be reasonably large. Nevertheless, the variables that
enter in the beta functions are then rather small because they involve a factor
of $1/(4\pi^2)$; $a_y(\mu) = \lambda(\mu) = 1/(4\pi)^2 = 0.6333 \times
10^{-2}$.  In the beta function $\beta_{a_y}$, the one-loop term $b_{a_y}^{(1)}
= 2.005 \times 10^{-4}$, and the two-loop term term $b_{a_y}^{(2)} = -0.4571
\times 10^{-5}$, so that the ratio of the two-loop to one-loop terms is
\beq
y=\lambda=1 \ \Rightarrow \ \frac{b_{a_y}^{(2)}}{b_{a_y}^{(1)}} = -0.02280 \ .
\label{beta_ay_two_over_one_loop}
\eeq
In the beta function
$\beta_{a_\lambda}$, the one-loop term $b_{a_\lambda}^{(1)} = 
1.043 \times 10^{-3}$ and $b_{a_\lambda} = -0.89135 \times 10^{-4}$, so that 
\beq
y=\lambda=1 \ \Rightarrow \ \frac{b_{a_\lambda}^{(2)}}{b_{a_\lambda}^{(1)}} 
= -0.0855 \ .  
\label{beta_alam_two_over_one_loop}
\eeq
We also note the values of the one-loop and two-loop beta functions for $a_y$
and $a_\lambda$:
\beq
y=\lambda=1\ \Rightarrow \ \frac{\beta_{a_y,1\ell}}{\beta_{a_\lambda,1\ell}} = 
\frac{b_{a_y}^{(1)}}{b_{a_\lambda}^{(1)}} = 0.1923
\eeq
and
\beq
y=\lambda=1\ \Rightarrow \ \frac{\beta_{a_y,2\ell}}{\beta_{a_\lambda,2\ell}} = 
\frac{b_{a_y}^{(1)}+b_{a_y}^{(2)}}
{b_{a_\lambda}^{(1)}+b_{a_\lambda}^{(2)}} = 0.2055 
\label{beta_ay_2loop_over_beta_alam_2loop}
\eeq
Thus, for this illustrative case with $y(\mu_0) = \lambda(\mu_0)=1$, the
two-loop term in $\beta_{a_y}$ makes only a small contribution relative to the
one-loop term, so that the perturbative expansion for $\beta_{a_y}$ is
reasonably reliable to this two-loop order, and similarly for
$\beta_{a_\lambda}$. 


\subsection{RG Flows}

To study the RG flows in this model, we begin by finding the fixed points, that
is the solutions to the simultaneous conditions $\beta_{a_y,n\ell}=0$,
$\beta_{a_\lambda,k\ell}=0$ for the values of loop orders $(n,k)$ that we
consider. We first note that the IR-free (trivial) fixed point
\beq
a_y^* = 0, \quad a_\lambda^* = 0\ , 
\label{ayalam_trivial_fp}
\eeq
is a solution to the beta functions for any of our $(n,k)$ cases.  Beyond this
IR-free fixed point, we find that the choice of loop order $(n,k)$ in the beta
functions is quite important for the appearance and location of fixed
points.  From Eqs. (\ref{bay_1loop})-(\ref{balam_2loop}), we calculate the
fixed point to be as follows:
\beq
{\rm case} \ (1,1) \ \Rightarrow \ {\rm no \ nonzero \ fixed \ points}. 
\label{su2xu1_11}
\eeq
\beq
{\rm case} \ (1,2)  \ \Rightarrow \ a_y^* = 0, \quad 
a_\lambda^* = \frac{1}{13}=0.07692.
\label{su2xu1_12}
\eeq
\beqs
{\rm case} \ (2,1)  \ & \Rightarrow \ & a_y^* = 
\frac{5}{318}(13\sqrt{13}-17)=0.4697, \cr\cr
& & a_\lambda^* = \frac{5}{638}(31-5\sqrt{13})=0.1020 \ . \cr\cr
& & 
\label{su2xu1_21}
\eeqs
\beqs
{\rm case} \ (2,2)  \ & \Rightarrow \ & {\rm two \ fixed \ points:} \cr\cr
 & & a_y^* = 0, \quad a_\lambda^* = \frac{1}{13}=0.07692 \ \ {\rm and} \cr\cr
 & & a_y^* = 0.4104, \quad a_\lambda^* = 0.1247 \ .
\label{su2xu1_22}
\eeqs

\begin{figure*}[hbt]
  \includegraphics[width=.9\textwidth]{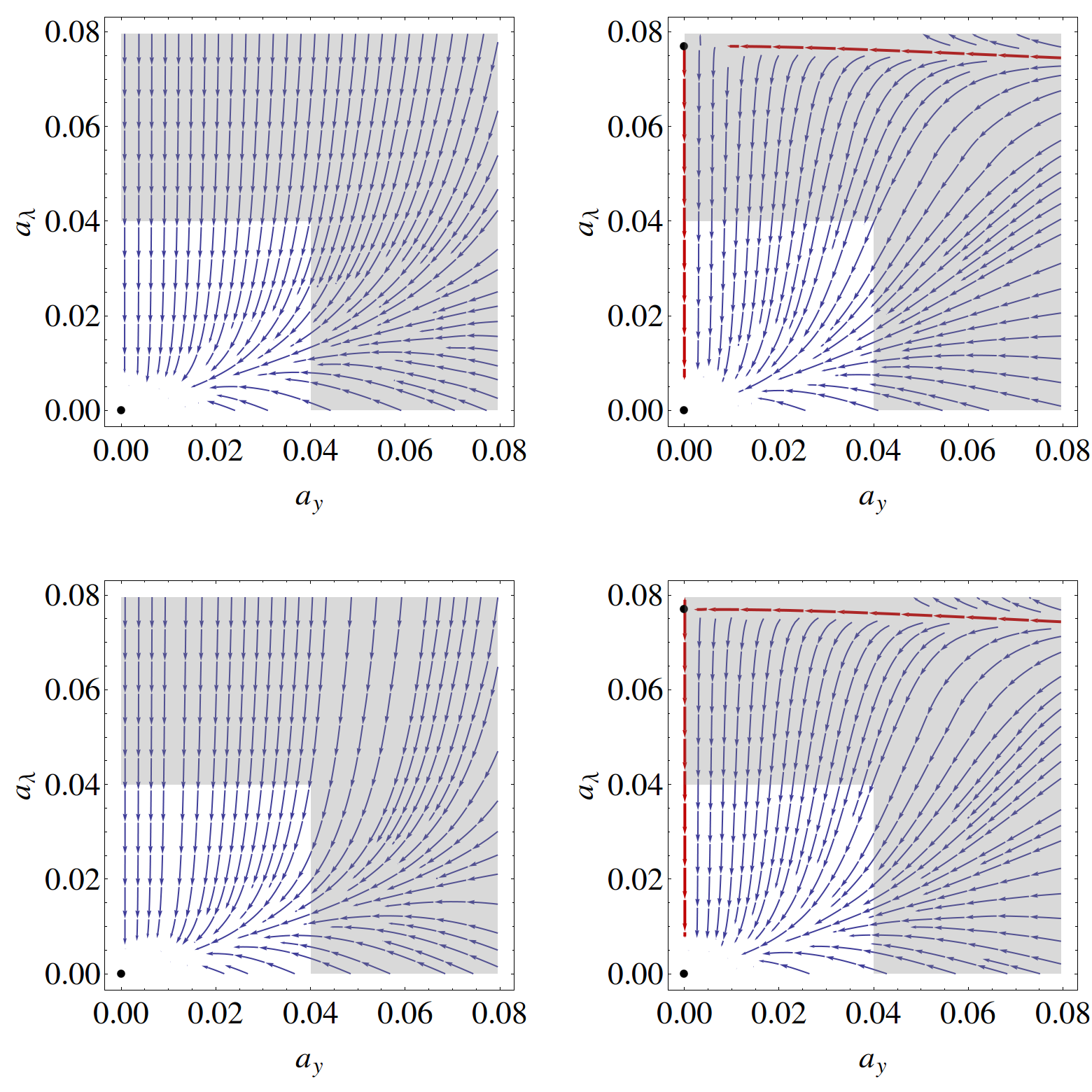}
  \caption{The renormalization-group flows for the
$\mathrm{SU}(2)\otimes\mathrm{U}(1)$ model with $0 \le a_y \le 1/(4\pi)$ and $0
\le a_\lambda \le 1/(4\pi)$. The white square region is where $0 \le a_y \le
0.04$ and $0 \le a_\lambda \le 0.04$, and the gray region occupies the rest of
the plot. The figures correspond to the following different choices of loop
order in the beta functions: (1,1) (upper left); (1,2) (upper right); (2,1)
(lower left); and (2,2) (lower right). The red flows for the cases (1,2) and
(2,2) originate along the eigendirections of the fixed points.}
  \label{fig:spLowNc}
\end{figure*}

The presence of a fixed point for such a low value of $a_\lambda$ as $1/13$
means that only a very small region of coupling space is independent of the
choice of $(n,k)$. In Fig. \ref{fig:spLowNc}, we see that the flows change
character based on $(n,k)$ when both $a_y$ and $a_\lambda$ are larger than
approximately 0.04. Note, in particular, that the plots where the two-loop term
$b_{a_\lambda}^{(2)}$ is included in $\beta_{a_\lambda}$ have concave flows
towards the trivial fixed point, whereas the ones where it is not have convex
flows towards the same in this region.

\begin{figure*}[hbt]
  \includegraphics[width=.9\textwidth]{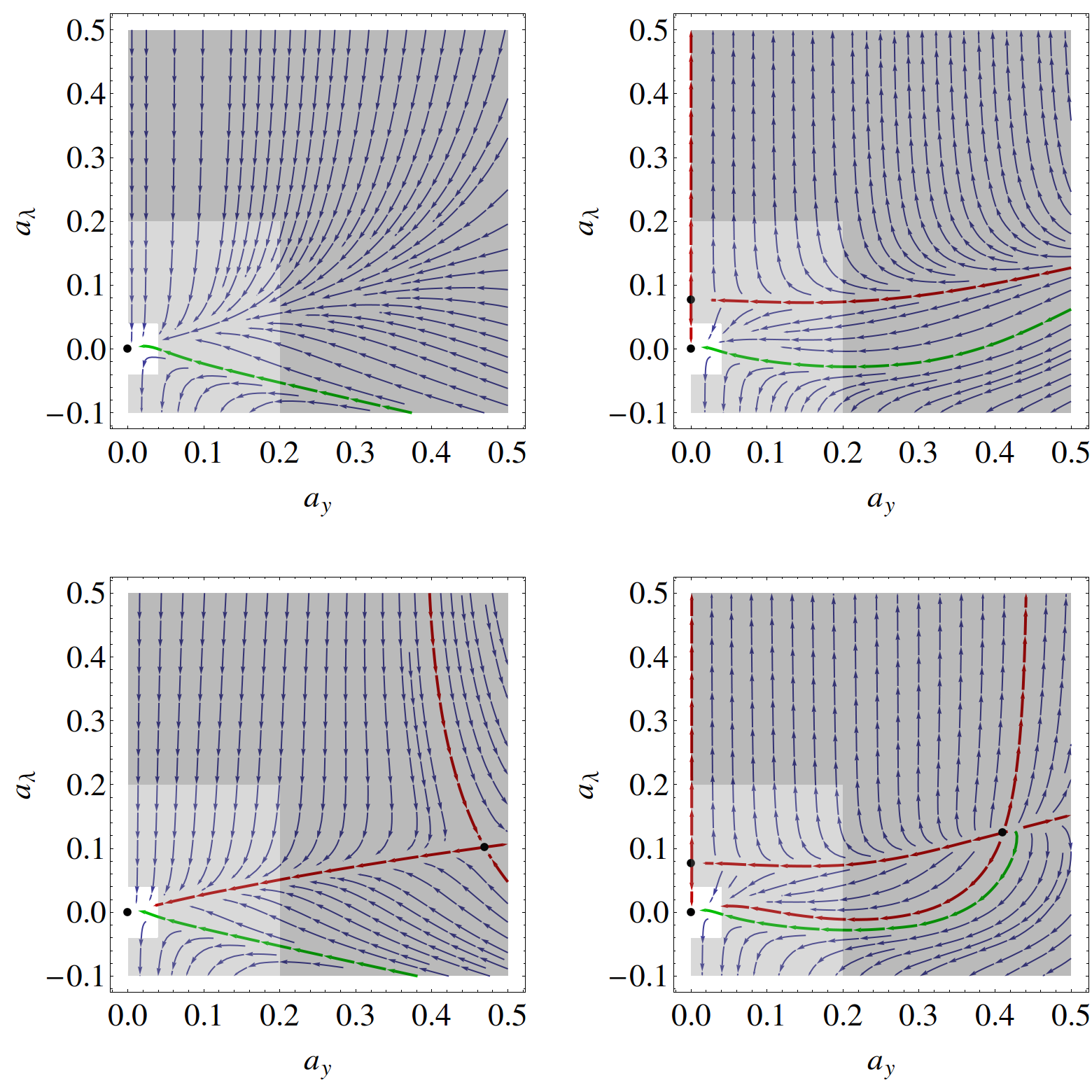}
  \caption{The renormalization-group flows for the 
$\mathrm{SU}(2)\otimes\mathrm{U}(1)$ model with $ 0 \le a_y \le 0.5$ and 
 $-0.1 \le a_\lambda \le 0.5$. The white region is 
  where $0 \le a_y \le 0.04$ and $-0.1 \le a_\lambda \le 0.04$; 
  the light gray region is where $0.04 \le a_y \le 0.2$ and $-0.1 \le a_\lambda \le
  0.2$; and the dark gray region occupies the rest of the figure. 
  The figures correspond to the following different choices of loop order in
  the beta functions: (1,1) (upper left); (1,2) (upper right); (2,1) 
 (lower left); and (2,2) (lower right). The green flows are the stable 
  manifolds in coupling constant space which bound the basins
  of attraction of the fixed point at the origin. The red flows in (1,2), 
  (2,1) and (2,2) originate along the eigendirections of the fixed points.}
  \label{fig:spHighNc}
\end{figure*}

If we let $a_y$ and $a_\lambda$ increase beyond $1/(4\pi)$, changes appear
quite rapidly (see Fig. \ref{fig:spHighNc}), which means that one cannot trust
the perturbative analysis to these orders in this region of couplings. With
this caveat in mind, we shall proceed to describe the RG flows. The first
striking difference is that if the two-loop term $\beta_{a_\lambda}^{(2)}$ in
the beta function $\beta_{a_\lambda}$ is included, then the flow ending in the
partially attractive fixed point at $a_y^* = 0, a_\lambda^* = 1/13$ is a
separatrix which divides a region where the flows end in the trivial fixed
point at the origin, from one where they increase to large values of
$a_\lambda$. The plots in this and the other figures were generated using the
Mathematica StreamPlot routine.  (Because the integration routine can lose some
numerical accuracy when the beta functions approach zero near fixed points, it
does not show arrows and associated RG flows very close to these fixed points.)

The second is that including the two-loop term in the Yukawa beta function
produces a fixed point where neither of the couplings is zero. However, the
impact that this has on the flow is very different in the (2,1) and (2,2)
cases. In the (2,1) case, the fixed point is partially attractive, and the flow
that reaches it from above forms a separatrix, separating a region where the
flows end at the origin from a region where they move toward larger values of $a_y$ in
the IR. In the
(2,2) case, the fixed point is totally repulsive, and the dominant term in the
beta functions is the $a_\lambda^3$ term in equation (\ref{balam_2loop}). This
term drives every flow, above the one originating in the eigendirection of
positive $a_y$ from the fixed point (marked in red on
Fig. \ref{fig:spHighNc}), towards larger $a_\lambda$ in the IR, which in turn
means that the dominant term in $\beta_{a_y,2\ell}$ will eventually be the
$a_ya_\lambda^2$ term, which drives $a_y\to0$ in the IR.

For the (2,2) flows that originate at the totally repulsive fixed point and go
in the direction of negative $a_\lambda$, there is a delicate balance between
terms driving them towards the origin and terms driving them towards
highly negative $a_\lambda$ in the IR. This balance is manifested in the stable
manifold (marked in green on Fig. \ref{fig:spHighNc}) which separates the
regions of convergence to the origin and flow to (unphysical) negative values. 


\section{${\rm SU}(N) \otimes {\rm SU}(N_f) \otimes {\rm U}(1)$ Model} 


\subsection{Field Content, Symmetry Group, and LNN Limit} 

In this section we study a model that is a two-fold generalization of the model
in the previous section.  First, we construct the model so that it is invariant
under a global symmetry group
\beq
G = {\rm SU}(N) \otimes {\rm SU}(N_f) \otimes {\rm U}(1) \ , 
\label{gglobal}
\eeq
rather than the ${\rm SU}(2) \otimes {\rm U}(1)$ group of the previous model.
We include an $N_f$-fold replication of the left-handed and right-handed
fermions.  The fermion content consists of (i) $\psi^a_{j,L}$, transforming as
a $(\fund,\fund)$ representation of ${\rm SU}(N) \otimes {\rm SU}(N_f)$, where
$a$ is an ${\rm SU}(N)$ group index taking on the values $a=1,...,N$, and $j$
is a copy (``flavor'') index, taking on the values $j=1,...,N_f$; and (ii)
$\chi_{j,R}$, with $j=1,...,N_f$, transforming as a $(1,\fund)$ representation
of ${\rm SU}(N) \otimes {\rm SU}(N_f)$.  The model also has a scalar field
$\phi^a$ transforming as a $(\fund,1)$ representation of ${\rm SU}(N) \otimes
{\rm SU}(N_f)$. The hypercharges are again taken to be nonzero and to satisfy
the conditions that $Y_\psi \ne Y_\chi$ and Eq. (\ref{hypercharge_rel}). The 
transformations of $\psi^a_{j,L}$ and $\chi_{j,R}$ under ${\rm SU}(N_f)$
are 
\beq
\psi^a_{j,L} \to \sum_{k=1}^{N_f} U_{jk} \psi^a_{k,L} \ , \quad 
\chi_{j,R} \to \sum_{k=1}^{N_f} U_{jk} \chi_{k,R} 
\label{psichitransf}
\eeq
where $U \in {\rm SU}(N_f)$.

The Lagrangian of this model is
\beqs
{\cal L} & = & \sum_{j=1}^{N_f} \Big [ \bar\psi_{j,L} i \dslash \psi_{j,L} + 
\bar\chi_{j,R} i \dslash \chi_{j,R} \Big ] \cr\cr
& - & y \sum_{j=1}^{N_f} [ \bar\psi_{j,L} \chi_{j,R} \phi + h.c. ] \cr\cr
& + & \partial_\mu \phi^\dagger \, \partial^\mu \phi - \mu_\phi^2 \phi^\dagger
\phi - \lambda (\phi^\dagger \phi)^2 \ , 
\label{lagrangian2}
\eeqs
where we have suppressed SU($N$) indices in the notation. 
The ${\rm SU}(N) \otimes {\rm U}(1)$ symmetry forbids the fermion
bilinears $\psi^{a \ T}_{j,L} C \psi^b_{k,L}$, $\chi_{j,R}^T C \chi_{k,R}$, and
$\bar \psi_{a,j,L}\chi_{k,R}$, so the fermions are massless.  Our requirement
of ${\rm SU}(N_f)$ invariance restricts the Yukawa coupling to the form given
in Eq. (\ref{lagrangian2}).  As before, we allow either sign of $\mu_\phi^2$
and impose the condition that $|\mu_\phi|$ be negligibly small relative to the
range of $\mu$ over which we calculate the RG flows (see also the end of Section \ref{sec:betaFuncs}). 

One of the motivations for this generalization is that it enables us to take
the combined limit
\beqs
& & N \to \infty \ , \quad N_f \to \infty  \ \ {\rm with} \ \ r \equiv 
\frac{N_f}{N} \ \ {\rm fixed}  \cr\cr
& & y \to 0 \ , \quad \lambda \to 0 \quad {\rm with} \  
\bar a_y \ {\rm and} \ \bar a_\lambda \ {\rm being} \cr\cr
& & {\rm finite \ functions \ of} \ \mu 
\label{lnn}
\eeqs
We will use the symbol $\lim_{LNN}$ for this limit, where ``LNN'' stands
for ``large $N_c$ and $N_f$'' 


\subsection{ Beta Functions}

To simplify the analysis, we take the LNN limit (\ref{lnn}).  In this limit,
from \cite{betafn} (see also \cite{betastability} we find 
\beq
b_{\bar a_y}^{(1)} = (1+2r)\bar a_y^2
\label{b_ay_1loop_lnn}
\eeq
\beq
b_{\bar a_y}^{(2)} = - 3r\bar a_y^3 
\label{b_ay_2loop_lnn}
\eeq
\beq 
b_{\bar a_\lambda}^{(1)} = 2(2 \bar a_\lambda^2 + 2 r \bar a_y \bar a_\lambda 
- r \bar a_y^2) 
\label{b_alam_1loop_lnn}
\eeq
and
\beq 
b_{\bar a_\lambda}^{(2)} = r \bar a_y (-8 \bar a_\lambda^2
-3 \bar a_y \bar a_\lambda + 2 \bar a_y^2) \ . 
\label{b_alam_2loop_lnn}
\eeq

\begin{figure}[hbt]
  \includegraphics[width=.9\columnwidth]{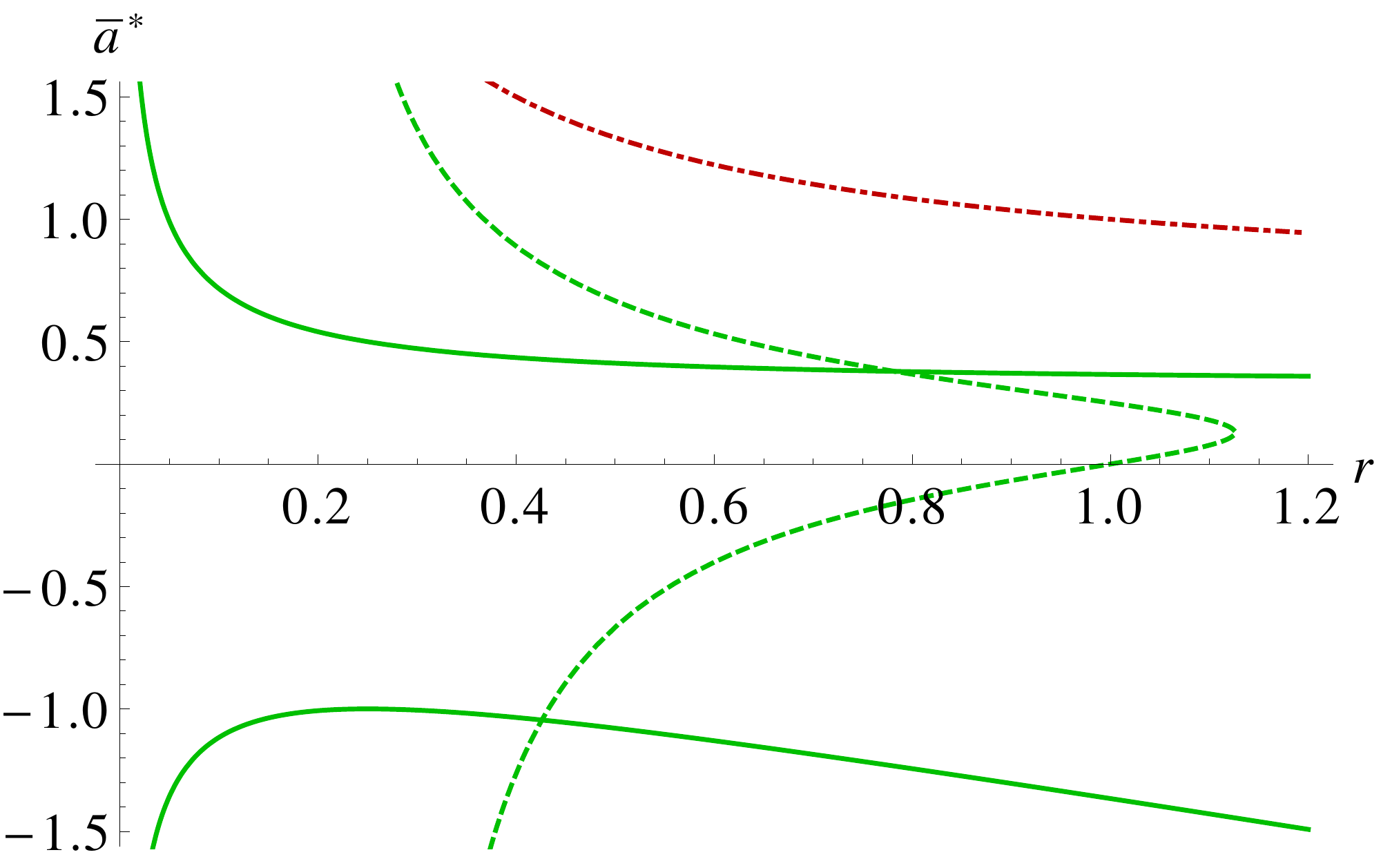}
  \caption{The fixed point values of (i) $\bar a_y$, denoted as $\bar a_y^*$
   and shown as the red, dot-dashed curve, and (ii) $\bar a_\lambda$, denoted
   as $\bar a_\lambda^*$ and shown as the green solid curve for the case (2,1)
   and green dashed curve for the (2,2) case, plotted as functions of $r=N_f/N$
   (with the LNN limit understood).  The curve for $\bar a_y^*$ is the same for
   the (2,1) and (2,2) cases, since, as discussed in the text, $\beta_{\bar
   a_y}$ is independent of $\bar a_\lambda$ to two-loop order.  The curves with
   $\bar a_\lambda$ negative are only formal, since the theory is unstable for
   $\bar a_\lambda \le 0$, i.e., $\lambda \le 0$.}
  \label{fig:bfps}
\end{figure}

We remark on some general properties of these terms.  First, because
$\beta_{\bar a_y}$ has an overall factor of $\bar a_y$, it follows that the
flow in $\bar a_y$ can never take $\bar a_y$ through zero to negative values of
$\bar a_y$.
For $y \ne 0$, the one-loop term in $\beta_{\bar a_y}$, namely $b_{\bar
a_y}^{(1)}$, is positive-definite and independent of $\bar a_\lambda$.  Hence,
provided that the initial values of $y$ and $\lambda$ at the starting point of
the integration are such that one can apply these perturbative calculations,
$\bar a_y$ decreases toward zero as $\mu$ decreases from the UV to the IR.
Since for $y \ne 0$, the two-loop term, $b_{\bar a_y}^{(2)}$, is negative, it
follows that the full two-loop beta function, $\beta_{\bar a_y,2\ell}= \bar
a_y^2[(1+2r)-3r\bar a_y]$ has a zero, which occurs at
\beq
\bar a_y^* = \frac{1+2r}{3r} \ , 
\label{betay_2loop_lnn_zero_aysol}
\eeq
independent of $\bar a_\lambda$.  For weaker Yukawa couplings, i.e., $\bar a_y
< \bar a_y^*$, $\beta_{\bar a_y,2\ell} > 0$, so the UV to IR flow is to still
weaker Yukawa couplings, while for $\bar a_y > \bar a_y^*$, $\beta_{\bar
a_y,2\ell} < 0$, so that the direction of the UV to IR flow is to larger $\bar
a_y$. Note that as $r$ decreases toward 0, $\bar a_y^*$ get sufficiently large
that one cannot trust the perturbative calculations, so this discussion is 
restricted to moderate values of $r$.  These results are shown in Fig.
\ref{fig:bfps}. 
For the range of $r$ shown in Fig. \ref{fig:bfps}, $\bar a_y^* \sim 1$. As is
evident from Eq. (\ref{betay_2loop_lnn_zero_aysol}), as $r \to \infty$, 
$\bar a_y$ approaches the limit 2/3 from above.

We next discuss the one-loop and two-loop terms in $\beta_{\bar
  a_\lambda}$. The analysis here is more complicated than that for $\beta_{\bar
 a_y}$, because whereas the one-loop and two-loop terms in $\beta_{\bar a_y}$
 depended only on $\bar a_y$, the one-loop and two-loop terms in $\beta_{\bar
 a_\lambda}$ depend on both $\bar a_\lambda$ and $\bar a_y$.  We find that the
one-loop term $b_{\bar a_\lambda}^{(1)}$ is positive (negative) if 
$\bar a_\lambda$ is larger (smaller) than the value 
\beq
\bar a_\lambda = \frac{1}{2} \Big [ -r + \sqrt{r(r+2)} \ \Big ] \bar a_y 
\label{b_alam_1loop_lnn_zero_alamsol}
\eeq
and zero if the equality in Eq. (\ref{b_alam_1loop_lnn_zero_alamsol})
holds. The condition in Eq. (\ref{b_alam_1loop_lnn_zero_alamsol}) is equivalent
to $\bar a_y = [1+\sqrt{1 + (2/r)}\ ] \bar a_\lambda$.  The solution for $\bar
a_\lambda$ in Eq. (\ref{b_alam_1loop_lnn_zero_alamsol}) is one of the two
solutions of the quadratic equation $b_{\bar a_\lambda}=0$; the solution with
the minus sign in front of the square root is unphysical because it leads to a
negative $\lambda$, and similarly in the equivalent solution for $\bar a_y$,
the other root with the minus sign in front of the square root is unphysical.
The fact that $b_{\bar a_\lambda}^{(1)} > 0$ for $\bar a_\lambda$ larger then
the value on the right-hand side of Eq. (\ref{b_alam_1loop_lnn_zero_alamsol}) means
that if the initial value of $\bar a_\lambda$ satisfies this condition, then in
the RG flow from the UV to the IR, $\bar a_\lambda$ decreases, and similarly,
if the initial value of $\bar a_\lambda$ is smaller than the value on the
right-hand side of Eq. (\ref{b_alam_1loop_lnn_zero_alamsol}), then $\bar a_\lambda$
increases in the RG flow from the UV to IR. 

We come next to the two-loop term in $\beta_{\bar a_\lambda}$, namely $b_{\bar
a_\lambda}^{(2)}$. Because this factorizes into a linear times a quadratic
factor in the LNN limit that we consider here, it is somewhat simpler to
analyze than $b_{a_\lambda}^{(2)}$ for the ${\rm SU}(2) \otimes {\rm U}(1)$
model.  We find that $b_{\bar a_\lambda}^{(2)}$ is negative (positive) if $\bar
a_\lambda$ is larger (smaller) than the value
\beq
\bar a_\lambda = \frac{1}{16} ( -3 + \sqrt{73} \ ) \,  \bar a_y = 
0.34650 \bar a_\lambda
\label{b_alam_2loop_lnn_zero_alamsol}
\eeq
(The solution of the quadratic with the opposite sign in front of the square
root is unphysical, since it renders $\lambda$ negative.)  The two-loop term
$b_{\bar a_\lambda}^{(2)}$ vanishes if $\bar a_y=0$ or if the condition in
Eq. (\ref{b_alam_2loop_lnn_zero_alamsol}) is satisfied.  Thus, for large $\bar
a_\lambda$ relative to $\bar a_y$, as least to the extent that our perturbative
calculations still apply, we thus find that the one-loop and two-loop terms in
the $\beta_{\bar a_\lambda,2\ell}$ have the opposite signs; $b_{\bar
a_\lambda}^{(1)} > 0$, while $b_{\bar a_\lambda}^{(2)} < 0$.  Similarly, for
sufficiently small $\bar a_\lambda$ relative to $\bar a_y$, these terms again
have opposite signs; $b_{\bar a_\lambda}^{(1)} < 0$, while $b_{\bar
a_\lambda}^{(2)} > 0$. It is thus plausible that the full two-loop $\beta_{\bar
a_\lambda,2\ell}$ would have a zero, where these terms cancel each other.

In Fig. \ref{fig:bfps} we show our solutions for the value of the fixed point
in the variable $\bar a_\lambda$ as a function of $r$. (Here and elsewhere, it
is implicitly understood that the LNN limit has been taken.)  The value of $r$
determines the value of the fixed point in $\bar a_y$, the existence or
non-existence of a fixed point in $\bar a_\lambda$, and, in the former case,
its value.  The solutions that yield a fixed point $\bar a_\lambda^*$ at
negative values are only formal, since the theory is unstable for $\bar
a_\lambda < 0$, i.e., $\lambda < 0$.  If $\bar a_\lambda$ is negative but
$|\bar a_\lambda|$ is sufficiently small, the theory may be metastable, but
considerations of metastability and estimates of tunneling times are beyond
the scope of our present analysis.  Thus, as regards $\bar a_\lambda$, there is
only a single physical fixed point, $\bar a_\lambda^*$, and the calculation for
the (2,1) case yields a value of $\bar a_\lambda^* \simeq 0.5$ in the range of
$r$ shown, for which perturbation theory may be reliable
down to $r \simeq 0.2$.  As $r \to \infty$, this curve for $\bar a_\lambda^*$
approaches the limit 1/3. For the (2,2) case, if $r < 1.0$, there is also only
one physical (positive) fixed point, $\bar a_\lambda^*$, but its value grows
more rapidly as $r$ decreases, so that one anticipates significant corrections
to the two-loop perturbative result already for $r$ decreasing below $r \simeq
0.4$. In the narrow interval of $r$ between $r=1.0$ and the value 
\beq
r^{(2,2)}_{merger} = \frac{31+12\sqrt{3}}{46} = 1.12575 
\label{r22merger}
\eeq
there are two physical fixed points for $\bar a_\lambda$.  We shall refer to
these as the upper and lower fixed points.  As $r$ increases through the value
$r^{(2,2)}_{merger}$, the upper and lower fixed points in $\bar a_\lambda$
merge and disappear.  This happens when the solution to the equation
$\beta_{\bar a_y,2\ell} = \beta_{\bar a_\lambda,2\ell} = 0$ becomes complex,
which happens at $r=r^{(2,2)}_{merger}$.  


\subsection{RG Flows}

\begin{figure*}[hbt]
  \includegraphics[width=.9\textwidth]{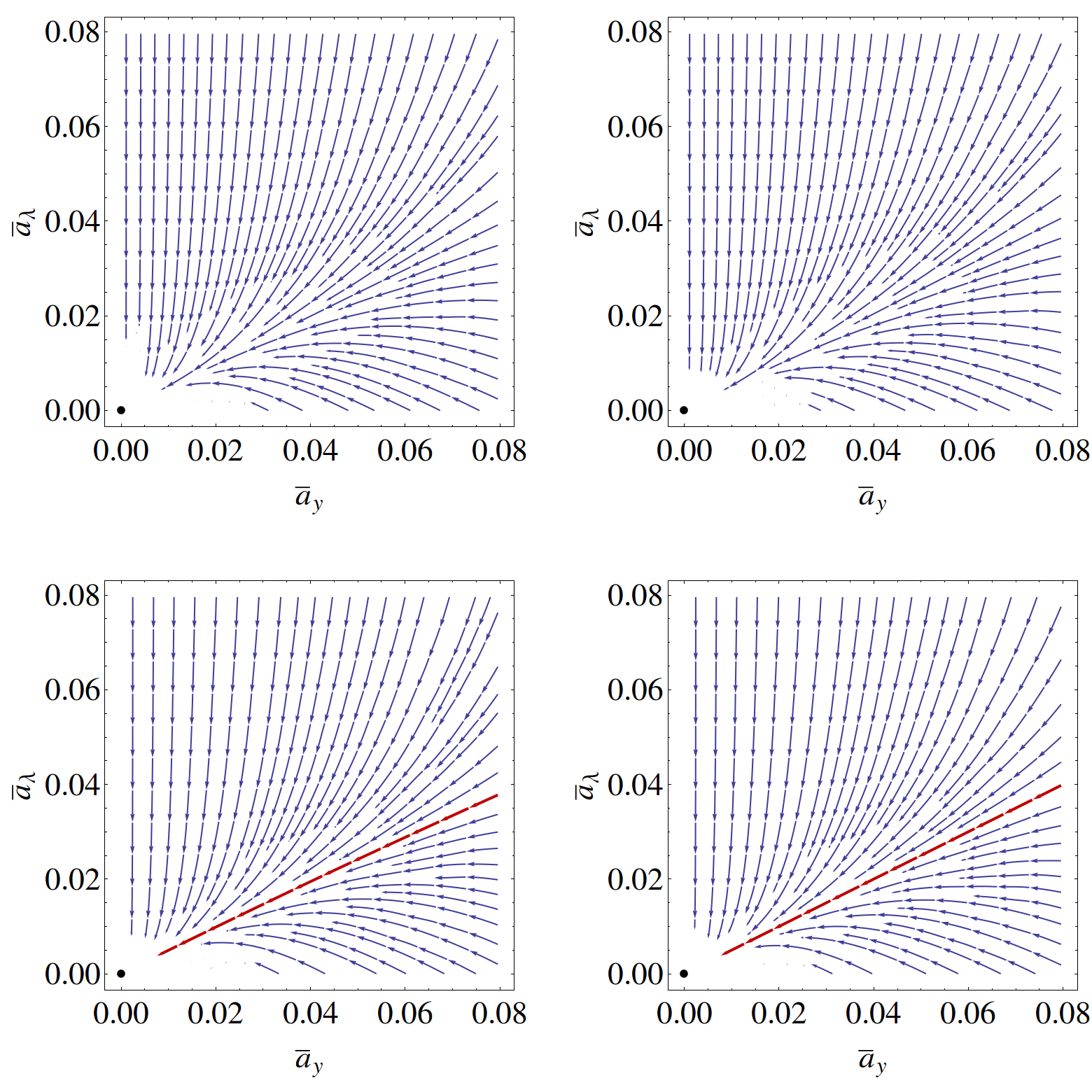}
  \caption{The renormalization-group flows for $r=0.5$ with $0 \le \bar a_y 
\le 1/(4\pi)$ and $0 \le \bar a_\lambda \le 1/(4\pi)$. The 
figures correspond to the following choices of inclusion of different-loop
terms in the beta functions: upper left: (1,1); upper right: (1,2); lower left:
(2,1); lower right: (2,2). The red flows in the (2,1) and (2,2) cases 
originate along the eigendirection of the upper fixed point 
 (see Figure \ref{fig:bfps}).}
  \label{fig:spLow05}
\end{figure*}

Here we present the results of our integration of the beta functions calculated
to various loop orders.  Our convention is to start the analysis at a 
high value of $\mu$ in the UV, integrate the renormalization-group equations
for $\bar a_y$ and $\bar a_\lambda$, and follow the flow from the UV to the
IR, and this is indicated by the direction of the arrows. In
Fig. \ref{fig:spLow05} we plot the RG flows for $r=0.5$ and
\beq
\bar a_y < \frac{1}{4\pi} \ , \quad \bar a_\lambda < \frac{1}{4\pi} \ ,
\ i.e., \ \frac{y^2 N}{4\pi} < 1 \ , \quad 
         \frac{\lambda N}{4\pi} < 1 \ . 
\label{ayalamrange}
\eeq
We find that for this value of $r$ and range of $\bar a_y$ and $\bar
a_\lambda$, the theory has only the IR fixed point at the 
IR-free point
\beq
(\bar a_y^*, \bar a_\lambda^*) = (0,0) \ . 
\label{irfp_zero}
\eeq
This can be understood as a result of the fact that the one-loop expression for
$\beta_{a_y}$, namely, $\beta_{a_y,1\ell}$, is positive and independent of
$a_\lambda$, so as $\mu$ decreases from the UV to the IR, $\bar a_y$ always
decreases.  Although the one-loop result for $\beta_{\bar a_\lambda}$, namely
$\beta_{\bar a_\lambda,1\ell}$, could initially be negative if the initial
value of $\bar a_y$ is such that $\bar a_y > (1+\sqrt{13} \ )\bar a_\lambda$,
as discussed above, $\beta_{\bar a_\lambda,1\ell}$ will eventually pass through
zero and become positive as $\bar a_y$ decreases through this zero, and as the
flow continues toward the IR thereafter, $\beta_{\bar a_\lambda,1\ell}$ will
remain positive.  This causes $\bar a_\lambda$ to vanish in the IR.

These results also provide an answer to a question that we posed at the
beginning, namely how robust the perturbative calculation of the RG flows are
to the inclusion of higher-loop terms in the beta function.  For this range
(\ref{ayalamrange}) of $\bar a_y$ and $\bar a_\lambda$, all four cases (1,1),
(1,2), (2,1), and (2,2) yield qualitatively similar flows. This serves as a
strong indication that for this range (\ref{ayalamrange}), our perturbative 
calculations are reliable.  

\begin{figure*}[hbt]
  \includegraphics[width=.9\textwidth]{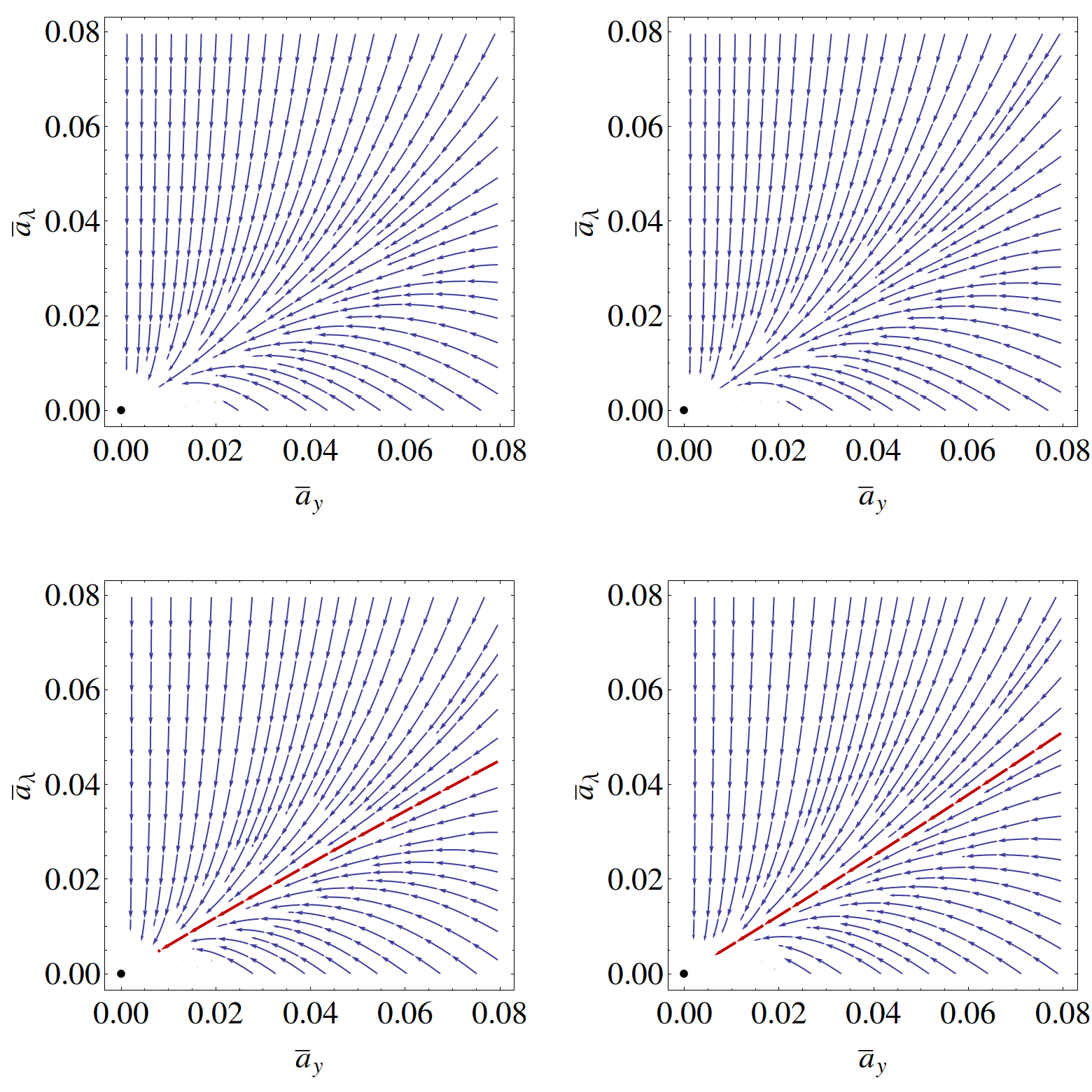}
  \caption{The renormalization-group flows for $r=1.1$ with $0 \le \bar a_y \le
   1/(4\pi)$ and $0 \le \bar a_\lambda \le 1/(4\pi)$. The figures correspond to
   the following choice of inclusion of different loop-order terms in the beta
   functions: upper left: (1,1); upper right: (1,2); lower left: (2,1); lower
   right: (2,2). The red flows in the (2,1) and (2,2) cases originate along the
   eigendirection of the upper fixed point (see Figure \ref{fig:bfps}).}
  \label{fig:spLow11}
\end{figure*}

Next, we increase $r$ from 0.5 to 1.1.  The results are shown in Fig. 
\ref{fig:spLow11}.  
We reach the same qualitative conclusions for this case $r=1.1$ as for
$r=0.5$.  

\begin{figure*}[hbt]
  \includegraphics[width=.9\textwidth]{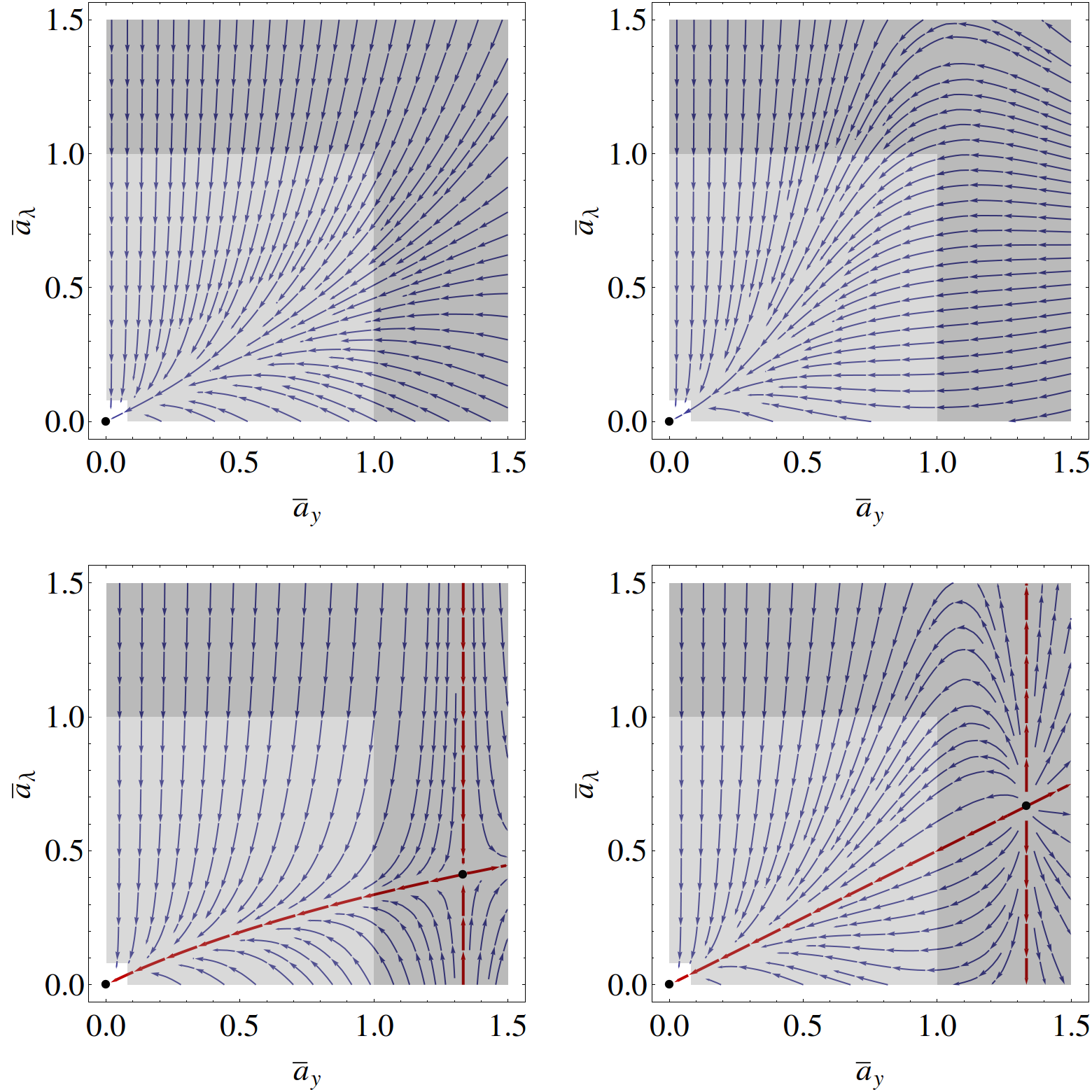}
  \caption{The renormalization-group flows for $r=0.5$ with $0 \le \bar a_y \le
  1.5$ and $0 \le \bar a_\lambda \le 1.5$. The white square region is where $0
  \le \bar a_y \le 1/(4\pi)$ and $0 \le \bar a_\lambda \le 1/(4\pi)$; the light
  gray region is where $1/(4\pi) \le \bar a_y \le 1$ and $1/(4\pi) \le \bar a_\lambda
  \le 1$; and the dark gray region occupies the rest of the plot.  The figures
  correspond to the following choices of inclusion of different loop-order
  terms in the beta functions: (1,1) (upper left); (1,2) (upper right); (2,1)
  (lower left); and (2,2) (lower right). The red flows in (2,1) and (2,2)
  originate along the eigendirections of the fixed points.}
  \label{fig:spHigh05}
\end{figure*}

\begin{figure*}[hbt]
  \includegraphics[width=.9\textwidth]{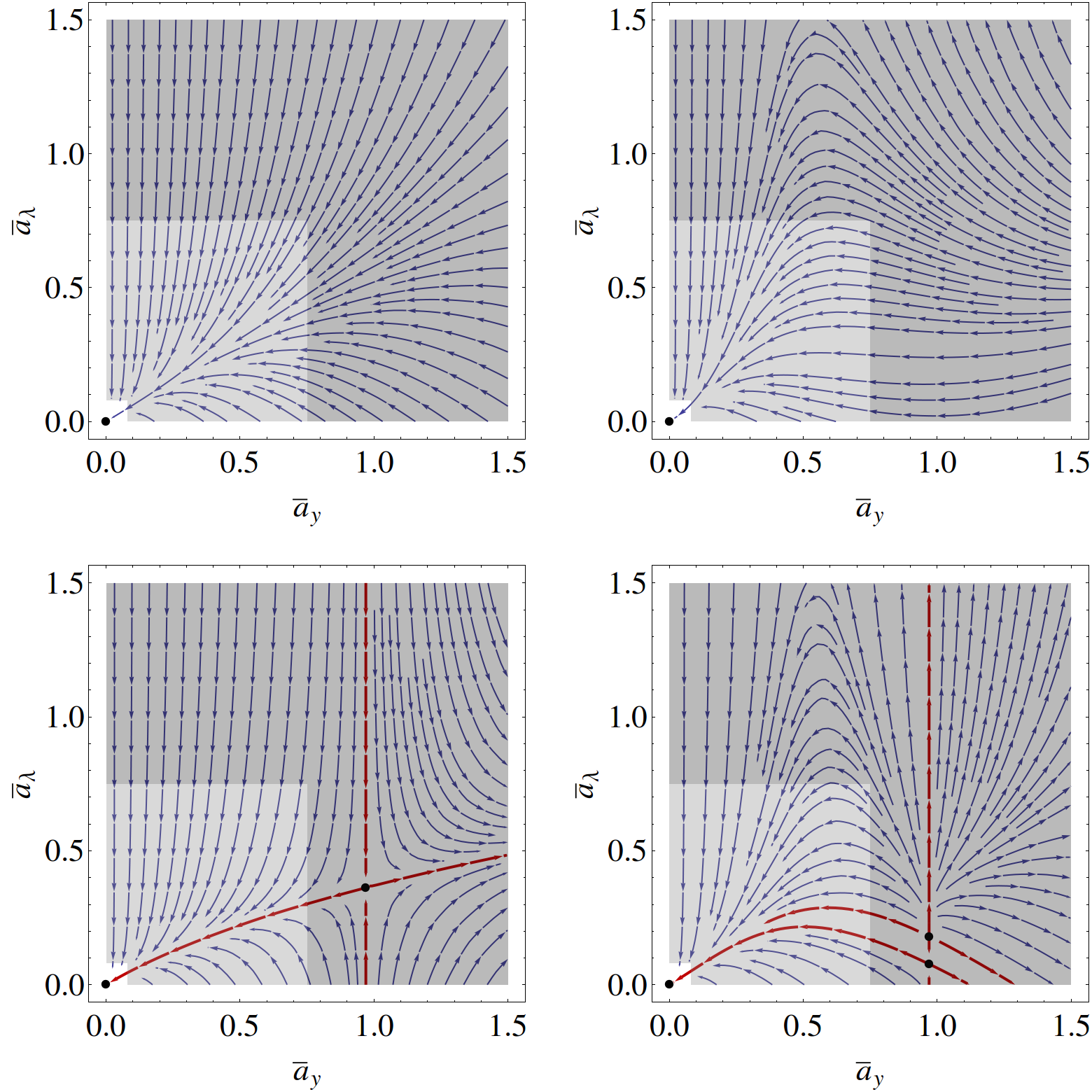}
  \caption{The renormalization-group flows for $r=1.1$ with $0 \le \bar
  a_y \le 1.5$ and $0 \le \bar a_\lambda \le 1.5$. The white square region is 
  where $0 \le \bar a_y \le 1/(4\pi)$ and 
        $0 \le \bar a_\lambda \le 1/(4\pi)$; the light gray region is were
  $1/(4\pi) \le \bar a_y \le 0.75$ and $1/(4\pi) \le \bar a_\lambda \le 0.75$; and 
 the dark gray occupies the rest of the plot. The figures correspond to the 
following choices of inclusion of different loop-order terms in the beta 
functions: (1,1) (upper left); (1,2) (upper right); (2,1) (lower left); and 
  (2,2) (lower right). The red flows in (2,1) and
  (2,2) originate along the eigendirections of the fixed points.}
  \label{fig:spHigh11}
\end{figure*}

We next study a larger range of $\bar a_y$ and $\bar a_\lambda$, namely 
$0 < \bar a_y < 1.5$ and $0 < \bar a_\lambda \le 1.5$.  We show the RG
flows for $r=0.5$ and $r=1.1$ in Figs. \ref{fig:spHigh05} and 
\ref{fig:spHigh11}. 

For reference, in these plots we distinguish three regions: (i) a white square
region where $0 \le \bar a_y \le 1/(4\pi)$ and $0 \le \bar a_\lambda \le
1/(4\pi)$; (ii) a light gray region where $1/(4\pi) \le \bar a_y \le 1$ and
$1/(4\pi) \le \bar a_\lambda \le 1$ ($1/(4\pi) \le \bar a_y \le 0.75$ and
$1/(4\pi) \le \bar a_\lambda \le 0.75$ in Fig. \ref{fig:spHigh11}); and (iii) a
dark gray region where $1 \le \bar a_y \le 1.5$ and $1 \le \bar a_\lambda \le
1.5$ ($0.75 \le \bar a_y \le 1.5$ and $0.75 \le \bar a_\lambda \le 1.5$ in
Fig. \ref{fig:spHigh11}). In the case where $r=0.5$ (Figure
\ref{fig:spHigh05}), the four light gray regions are still quite similar, but
now the inclusion of the two-loop term in $\beta_{\bar a_\lambda}$ has a
significant effect.  In the left-hand plots where this term is not included, we
note that the flows that reach the fixed points seem to be attracted to a
central flow, which, in the (2,1) (lower left) plot is identified with the one
flowing in the eigendirection from the upper fixed point.  In the right-hand
plots that include the two-loop term in $\beta_{\bar a_\lambda}$, this behavior
is reversed for relatively large values of $\bar a_y$.  In (1,1) and (2,1)
cases, the RG flows in the light gray region where $\bar a_y \le 1$ and $\bar
a_\lambda\le1$, look similar to the flows in the white square region where
$\bar a_y \le 1/(4\pi)$ and $\bar a_\lambda \le 1/(4\pi)$.

The largest changes in the flows occur in the dark gray area where $\bar a_y$
and $\bar a_\lambda$ are largest. When considering this region, it is
important to recall that this is where we expect perturbation theory to break
down, partly because higher-order terms in the beta functions are of comparable
size compared with lower-order terms, and partly because completely
nonperturbative effects such as fermion condensates can appear
for such strong values of the couplings.  However, continuing in the context of
the perturbative analysis, we see that fixed points appear in the (2,1) and 
(2,2) plots, and correspondingly the flows are changed by their presence. 

The inclusion of the two-loop term in $\beta_{\bar a_\lambda}$ fundamentally
changes the nature of the fixed points. In the (2,1) plot, we see that the
non-trivial fixed point is attractive along the vertical direction, and
repulsive along the approximately horizontal direction, but the fixed point in
the (2,2) case occurs at a roughly similar position, it is now repulsive in all
directions.

In Fig. \ref{fig:spHigh11}, we note that (1,1), (1,2), and (2,1) plots are
similar to those in Fig. \ref{fig:spHigh05}, except that the fixed point in the
(2,1) plot now occurs at a value of $\bar a_y<1$.  However, in the (2,2) plot,
the flows are very different. Most dramatically, the lower fixed point (see
Fig. \ref{fig:bfps}) has become positive, and is very close to merging with the
upper one.

Thus, our comparative calculations of RG flows for these (1,1), (1,2), (2,1),
and (2,2) cases in this model show that a perturbative calculation of the RG
flows and fixed points is reasonably reliable for the region $0 \le \bar a_y
\lsim 1/(4\pi)$ and $0 < \bar a_\lambda \lsim 1/(4\pi)$ but is unreliable when
these variables increase to sizes of order 1 or greater.

\section{Conclusions}

In summary, in this paper we have calculated renormalization-group flows and
resultant fixed points in scalar-fermion theories depending on two couplings, a
Yukawa coupling $y$ and a quartic scalar self-coupling $\lambda$.  We have
addressed a fundamental question pertaining to the RG flows in these theories,
namely the question of the range of values of $y$ and $\lambda$ for which these
flows can be determined reliably using the beta functions $\beta_y$ and
$\beta_\lambda$ calculated up to various respective loop orders. To investigate
this, we have focused on two models and have calculated these flows
using the $n$-loop beta function $\beta_{y,n\ell}$ and the $k$-loop beta
function $\beta_{\lambda,k\ell}$ with $(n,k)=(1,1)$, (1,2), (2,1), (2,2).  We
have presented our results in a set of convenient variables, $a_y$ and
$a_\lambda$ for a model with a global ${\rm SU}(2) \otimes {\rm U}(1)$ symmetry
group and $\bar a_y$ and $\bar a_\lambda$ in the limit (\ref{lnn}) of a model
with a ${\rm SU}(N) \otimes {\rm SU}(N_f) \otimes {\rm U}(1)$ global symmetry
group.  Our results provide a quantitative answer to this question. In future
work, it would be worthwhile to extend the perturbative calculations of the
beta functions to higher loop orders and to investigate connections between
semiperturbative properties at moderately strong coupling and nonperturbative
phenomena in the scalar and fermion sectors of the models.


\begin{acknowledgments}

This research was partly supported by the Centre for Particle Physics
Phenomenology-Origins (CP3-Origins) of the Southern Denmark University which
is partially funded by the Danish National Research Foundation, grant number
DNRF90 (E.M.) and by the NSF grant NSF-PHY-13-16617 (R.S.).

\end{acknowledgments}


\newpage



\begin{thebibliography}{99}
\bibitem{mphi}
%
The purpose of this assumption of negligibly small $m_\phi$ compared with the
range of $\mu$ of interest for our RG flows is to ensure that the $\phi$ field
is dynamical; if $m_\phi$ were $>> \mu$ for the values of $\mu$ of interest,
then we could integrate it out, obtaining a low-energy effective field theory
consisting of just the fermions $\psi$ and $\chi$ with a resultant four-fermion
operator $\propto (1/m_\phi^2)\sum_a [\bar\psi_{a,L}\chi_R][\bar\chi_R
\psi^a_L] + h.c.$.
\end{thebibliography}
\end{document}